# Complexity Analysis Approach for Prefabricated Construction Products Using Uncertain Data Clustering

Wenying Ji[1], S.M.ASCE; Simaan M. AbouRizk[2], M.ASCE; Osmar R. Zaïane[3]; Yitong Li[4], S.M.ASCE

[1]Ph.D. Candidate, University of Alberta, Department of Civil & Environmental Engineering, 9105 116 ST, 5-080 NREF, Edmonton, Alberta, T6G 2W2, Canada, wenying.ji@ualberta.ca

[2]Professor, University of Alberta, Department of Civil & Environmental Engineering, 9105 116 ST, 5-080 NREF, Edmonton, Alberta, T6G 2W2, Canada, abourizk@ualberta.ca, Corresponding Author

[3]Professor, University of Alberta, Department of Computing Science, 4-43 Athabasca Hall, Edmonton, Alberta, T6G 2E8, Canada, zaiane@ualberta.ca

[4]Undergraduate Student, University of Alberta, Department of Civil & Environmental Engineering, 9105 116 ST, 5-080 NREF, Edmonton, Alberta, T6G 2W2, Canada, yitong2@ualberta.ca

## ABSTRACT

This paper proposes an uncertain data clustering approach to quantitatively analyze the complexity of prefabricated construction components through the integration of quality performance-based measures with associated engineering design information. The proposed model is constructed in three steps, which (1) measure prefabricated construction product complexity (hereafter referred to as product complexity) by introducing a Bayesian-based nonconforming quality performance indicator; (2) score each type of product complexity by developing a Hellinger distance-based distribution similarity measurement; and (3) cluster products into homogeneous complexity groups by using the agglomerative hierarchical clustering technique. An illustrative example is provided to demonstrate the proposed approach, and a case study of an industrial company in Edmonton, Canada, is conducted to validate the feasibility and applicability of the proposed model. This research inventively defines and investigates product complexity from the perspective of product quality performance with design information associated. The research outcomes provide simplified, interpretable, and informative insights for practitioners to better analyze and manage product complexity. In addition to this practical contribution, a novel hierarchical clustering technique is devised. This technique is capable of clustering uncertain data (i.e., beta distributions) with lower computational complexity and has the potential to be generalized to cluster all types of uncertain data.

**INTRODUCTION**

As the implementation of modular construction expands, an increasing number of prefabricated construction products are being engineered and manufactured in fabrication shops. Construction products are heterogeneous in nature and are characterized by various combinations of design attributes, which, in turn, impacts the complexity involved in producing or assembling these products. As product complexity increases, so too do the skills, knowledge, management efforts (e.g. training and quality control), and resource support (e.g. specialized tools and technologies) required for successful performance. Inadequate management of product complexity, therefore, can result in cost and schedule overruns and can hamper overall project delivery.

The heterogeneous nature of construction product design, together with various levels of production knowledge and skill, has made the quantification of product complexity difficult in practice. In recent years, researchers have successfully correlated product complexity with product quality performance in the manufacturing industry, indicating that, for practical purposes, product quality performance can be used as an indicator or surrogate marker of product complexity (Antani 2014; Novak and Eppinger 2001; Williams 1999). Recently, analytically-based quality management systems, which facilitate quantitative quality performance measurements with design information associated, have been developed (Ji and AbouRizk 2017a; Ji and AbouRizk 2017b). Integrating these systems to generate a single indicator of product quality performance from which product complexity can be inferred would alleviate the need of practitioners to perform detailed, time-consuming analyses of complex, unreliable, subjective factors (e.g. design information and operator knowledge and skill). A quality performance-based product complexity indicator, however, has yet to be defined or developed within the construction domain.

The aim of the present study is to develop, validate, and implement an uncertain data clustering approach that is capable of quantifying and clustering quality performance-based product complexity indicators (hereafter referred to as product complexity indicators) from quality management and engineering design information. Specifically, the proposed approach has developed a framework that can provide (1) accurate and reliable measurements of product complexity indicator uncertainty; (2) meaningful assessments of product complexity indicator distribution similarity; and (3) an interpretable clustering of products with similar complexity indicators. The content of this paper is organized as follows: First, a comprehensive literature review is provided to demonstrate the rationale of the proposed research. Then, details of the methodology are introduced. To elaborate on the implementation of the proposed approach, an illustrative example is provided. Finally, the feasibility and applicability of the proposed approach are validated following a practical case study of industrial pipe weld complexity analysis. In addition to providing simplified, interpretable, and informative insights for understanding construction product complexity using quality management and engineering design information, this research also develops a novel Hellinger distance-based hierarchical clustering technique for grouping uncertain data (i.e., probability distributions).

**RATIONALE**
**Product Complexity**
In the construction research domain, construction project complexity has been primarily investigated from four perspectives: (1) influencing factors contributing to project complexity; (2) the impact of project complexity; (3) project complexity measurement methods; and (4) management of project complexity (Luo et al. 2017). Throughout these studies, product complexity has been found to influence overall project complexity (Baccarini 1996; Senescu et al. 2012; Williams 1999). In spite of these findings, product complexity has not been conceptually defined and thoroughly analyzed within construction management literature. In this study, the authors define product complexity as:

*"The level of constructing difficulty based on the product's design and on the knowledge and ability of an operator to construct a product given its specific design information."*

This definition is consistent with informal statements in product design and development literature (Baldwin and Clark 2000; Galvin and Morkel 2001; Novak and Eppinger 2001). For instance, Novak and Eppinger (2001) stated that "[t]he effect of this product design choice on the outsourcing decision can be profound, as greater product complexity gives rise to coordination challenges during product development."

Interviews conducted with five industrial construction companies in Edmonton, Canada, highlighted the difficulty that these organization have with determining product complexity from design information. Currently, complexity is assessed by examining the detailed design information of each product type. Given that there may be hundreds of product types in a single project, establishing product complexity is often a costly, time-intensive endeavor. Practitioners would benefit from the development of a framework that could rapidly generate a simple, reliable indicator of product complexity for estimating purposes. Several researchers in manufacturing literature have indicated that complexity can be reliably estimated from quality performance data (Antani 2014; Novak and Eppinger 2001).

While quality performance data are captured in practice, prefabricated products are often inspected as either conforming or nonconforming to specified quality standards and quality performance data cannot, therefore, be represented numerically (Ji and AbouRizk 2017a). Research conducted by Ji and AbouRizk (2017a) has quantitatively solved this issue by providing a Bayesian statistics-based analytical solution (i.e., a beta distribution) to estimate fraction nonconforming performance uncertainty at a product-level through the investigation of both quality management and engineering design information. The issue of assessing product complexity indicator in construction is further complicated by the myriad of prefabricated products that may be involved during project delivery. Although complexity of each product may be quantified, these data must be reduced into a format that is simple, interpretable, and informative for industry professionals (e.g. design and operations personnel). Notably, however, the uncertain nature of product complexity renders traditional clustering methods inappropriate for solving uncertain data clustering problems. A method capable of rapidly and reliably estimating product complexity and clustering hundreds of products of similar complexity into a manageable number of classification groups would improve the practice of product complexity analysis and management.

## Uncertain Data Clustering

In data mining, cluster analysis or clustering is the process of partitioning a set of objects in such a way that objects in a cluster are more similar to one another than to the objects in other clusters. An advantage of data clustering is that clustering can, automatically, lead to the discovery of previously unknown groups within data. Clustering as a standalone tool can be implemented to gain insights into the distribution of data and to observe the characteristics of each cluster (Han et al. 2011).

For many application domains, the ability to unearth valuable knowledge from a dataset is impaired by unreliable, erroneous, obsolete, imprecise, and noisy data (Schubert et al. 2015; Züfle et al. 2014)—or, in other words, uncertain data that is commonly described by a probability distribution (Jiang et al. 2013; Pei et al. 2007). Uncertain data are found in modeling situations where a mathematical model only approximates the actual nonconforming quality control process. Clustering uncertain data (i.e., probability distributions) is associated with substantial challenges concerning modeling similarity between uncertain objects and regarding the development of efficient computational methods (Jiang et al. 2013). Traditional clustering methods, such as partitioning-based clustering methods (e.g. k-means) and density-based clustering methods (e.g. DBSCAN), are dependent on geometric distances (e.g. Euclidean distance and Manhattan distance) between observations (Han et al. 2011). Such distances are not capable of grouping uncertain objects that are geometrically indistinguishable, such as products with similar repair rates that vary in terms of quality performance.

Jiang et al. (2013) were the first to use Kullback-Leibler (KL) divergence, which is a special type of $f$-divergence to measure distribution similarity, for uncertain data clustering problems. However, computing KL divergence to measure the similarity between complex distributions is very time-consuming and may even be infeasible (Jiang et al. 2013). The derivation of KL divergence between two beta distributions involves calculations of complicated digamma functions, thereby requiring additional computational efforts.

The Hellinger distance is another type of $f$-divergence that is widely used to quantify the similarity between two probability distributions in the field of statistics. The Hellinger distance, however, has yet to be used for solving uncertain data clustering problems in the data mining domain. In contrast to KL divergence, an analytical solution for measuring the similarity of beta distributions, which largely reduces the computational complexity of uncertain data clustering problems, exists. Therefore, the Hellinger distance is used in this research to model the similarity between distributions for product complexity clustering purposes. The mathematical proof is provided in Appendix 1. Notably, the Hellinger distance-based uncertain data clustering method proposed here can be further generalized to other types of uncertain data (i.e., probability distributions).

## METHODOLOGY

The proposed methodology is conducted following three steps. First, to measure the prefabricated construction product complexity indicator, a Bayesian statistics-based quality performance measurement (i.e., a posterior distribution of fraction nonconforming), which incorporates uncertainty, is introduced. Second, to develop a systematic product complexity indicator scoring approach, the Hellinger distance is used to measure complexity indicator similarity between various types of products. Finally, to cluster product complexity indicators into homogeneous groups, the agglomerative hierarchical clustering technique is adopted using the obtained Hellinger distance-based similarity measure. Details of the systematic and theoretical analysis of these steps are discussed as follows.

***Step 1. Quality Performance-based Product Complexity Indicator***

To quantitatively measure the product complexity indicator, a quality performance measurement termed fraction nonconforming, which represents the ratio of the number of nonconforming items $X$ in the sample to the sample size $n$, is utilized. Fraction nonconforming can be mathematically expressed as Eq. (1) (Montgomery 2007).

$$\hat{p} = \frac{X}{n} \qquad (1)$$

To appropriately incorporate the sampling uncertainty of the population fraction nonconforming variable $p$ when data are obtained from a sample, a Bayesian statistics-based analytical solution has been developed to determine the posterior distribution of the fraction nonconforming $p$ (Ji and AbouRizk 2017a). The posterior distibution uses a non-informative prior distribution $Beta(1/2,\ 1/2)$. It is given as Eq. (2).

$$P(p|X) = Beta(X + 1/2, n - X + 1/2) \qquad (2)$$

This Bayesian statistics-based solution, which is capable of updating the posterior distribution by combining previous knowledge and real-time data, has been demonstrated to be more accurate, reliable, and interpretable than the traditional statistical methods (Ji and AbouRizk 2017a).

As discussed previously, product complexity has been found to be positively correlated with product quality performance. In this research, the posterior distribution of fraction nonconforming $p$ is used to assess the product complexity indicator, termed $Cplx$. Therefore, the posterior distribution of $Cplx$ is identical to the posterior distribution of the fraction nonconforming, as shown in Eq. (3).

$$P(Cplx|X) = P(p|X) = Beta(X + 1/2, n - X + 1/2) \qquad (3)$$

This posterior distribution measures the product complexity indicator for a certain type of construction product. In the following step, the complexity indicator distribution similarity measurement and the complexity indicator scoring approach are introduced to evaluate product complexity in a systematic and interpretable way.

***Step 2. Product Complexity Indicator Scoring***

In this step, the product complexity indicator is scored by accounting for uncertainty. The Hellinger distance is introduced to measure the distribution similarity of product complexity indicators. The distances obtained for paired products are used to construct a Hellinger distance matrix, which is required for product complexity indicator clustering as follows.

To score the product complexity indicator, the distribution similarity of product complexity indicators between all types of products should be measured. A significant challenge in modeling distribution similarity is that the distribution similarity cannot be captured by geometric distances, such as the Euclidean distance or the Manhattan distance. In statistics, $f$-divergence is a function, $D_f(P||Q)$, that measures the similarity between two probability distributions (Liese and Vajda 2006). The Hellinger distance is a special case of $f$-divergence, which was defined in terms of the Hellinger integral by Ernst Hellinger in 1909 (Hellinger 1909). The reason for choosing Hellinger distance is that, for measuring the similarity between two beta distributions, the Hellinger distance has a closed-form solution, which largely reduces the computational efforts compared to other cases of $f$-divergences.

Conceptually, the Hellinger distance between two distributions, $P = \{p_i\}$ i $\in$ [n] and $Q = \{q_i\}$ i $\in$ [n], is defined as Eq. (4) (Hellinger 1909).

$$H(P, Q) = \frac{1}{\sqrt{2}} \left\| \sqrt{P} - \sqrt{Q} \right\|_2 = \frac{1}{\sqrt{2}} \sqrt{\int (\sqrt{p_i} - \sqrt{q_i})^2} \tag{4}$$

To measure the product complexity indicator similarity, the specialized Hellinger distance between two beta distributions, $X_t \sim Beta(a_1, b_1)$ and $Y_t \sim Beta(a_2, b_2)$, is derived as Eq. (5).

$$H(X_t, Y_t) = \sqrt{1 - \frac{Beta(\dfrac{a_1 + a_2}{2}, \dfrac{b_1 + b_2}{2})}{\sqrt{Beta(a_1, b_1) \times Beta(a_2, b_2)}}} \tag{5}$$

Where $0 < H(X_t, Y_t) < 1$. The Hellinger distance represents the similarity measurement between two product complexity indicator distributions: the larger the distance, the smaller the similarity between the distributions. A detailed mathematical proof for the closed-form solution is provided in Appendix 1.

Using the calculated Hellinger distances for all pairs of products, a two-dimensional distance matrix is constructed. The obtained Hellinger distance matrix $[M = (x_{ij})$ with $1 \leq i, j \leq N]$ is a distance matrix containing all complexity indicator similarity measurements. The entry $x_{ij}$ represents the similarity measurement between product types $i$ and $j$. The obtained matrix always adheres to the following properties: (1) the entries on the main diagonal are all zero (i.e., $x_{ij} = 0$ for all $1 \leq i \leq N$); (2) all the off-diagonal entries are in the range of 0 to 1 (i.e., $0 \leq x_{ij} \leq 1$ $if$ $i \neq j$); and (3) the matrix is symmetric (i.e., $x_{ij} = x_{ji}$).

This Hellinger distance matrix, however, can only demonstrate the quantitative distribution similarity measurements for each pair of product types. To determine the sequence of the complexity indicator scores of all product types, medians of the posterior distributions of $Cplx_i$ are compared. $P(0.5, Cplx_i|X_i)$ represents the 50% quantile (i.e., median) of the $Cplx$ distribtution. Therefore, the most non-complex product can be searched by indexing $Min(P(0.5, Cplx_i|X_i))$. If multiple distributions possess the same median, the distribution with the smaller variation is considered the less complex product.

Here, it is assumed that the ascendingly sorted product complexity indicator scores follow the sequence $(Cplx\ Score_n)_{n \in N}$, which denotes a sequence whose $nth$ element is given by the variable $Cplx\ Score_n$. $P_n$ is the probability distribution of the $nth$ scored product complexity indicator in the sequence $(Cplx\ Score_n)_{n \in N}$, and the sequence of $Cplx\ Score_n$ is defined by the recurrence relation expressed as Eq. (6):

$$Cplx\ Score_n = Cplx\ Score_{n-1} + H(P_n, P_{n-1})$$

With seed value $Cplx\ Score_1 = 0$

(6)

Where, $H(P_n, P_{n-1})$ represents the Hellinger distance between the $nth$ and $(n-1)th$ distributions of the sequenced product complexity indicators. Explicitly, the recurrence yields the following equations:

$$Cplx\ Score_2 = Cplx\ Score_1 + H(P_2, P_1)$$
$$Cplx\ Score_3 = Cplx\ Score_2 + H(P_3, P_2)$$
$$Cplx\ Score_4 = Cplx\ Score_3 + H(P_4, P_3)$$
$$\dots$$

(7)

All the involved Hellinger distances are available and can be indexed from the obtained Hellinger distance matrix. By using the recurrence relation defined as Eq. (6), complexity indicator scores for all types of products can be calculated. After all complex scores are derived, they are scaled to a range from 0 to 10, where a score of 10 represents the most complex product.

While the complexity scoring is used for clustering purposes, it also has a practical benefit. Transformation of uncertain quality performance distributions (i.e., beta distributions) into deterministic numbers, ranging from 0 to 10, can also reduce the interpretation load of practitioners, particularly for non-quality associated industrial personnel.

***Step 3. Product Complexity Indicator Clustering***

A method capable of clustering products of similar complexity indicators would improve product analysis and management, especially when a vast number of product types are involved. A useful summarization tool, which provides an interpretable visualization of the data, is, therefore, needed. Among multiple clustering techniques, hierarchical clustering is selected due to its ease of use and to the interpretability of the results. In addition, compared with partitioning-based clustering methods (e.g., k-means) and density-based clustering methods (e.g., DBSCAN), hierarchical clustering avoids treating data as an outlier. This characteristic is desired in this product complexity clustering problem because each type of product should be clustered into a complexity group rather than be excluded as an outlier. In data mining, hierarchical clustering is a method of cluster analysis that works by grouping similar data objects into a hierarchy or "tree" of clusters (Han et al. 2011). Visualizing this hierarchy provides a useful visual summary of the data. The agglomerative hierarchical clustering method begins by treating each object as an individual cluster and then iteratively merging clusters into larger and larger clusters until all objects are merged into a single cluster. To determine which clusters should be combined, a measure of similarity between sets of clusters is required. This is achieved by selecting an appropriate metric—in this case, the Hellinger distance—and a complete-linkage criterion that specifies the similarity of clusters as a function of the pairwise distances of the observations within the clusters. The complete-linkage criterion considers the distance between two clusters to be equal to the largest distance from any member of one cluster to any member of the other cluster. Complete-linkage tends to find compact clusters of approximately equal diameters and achieves more accurate clustering results.

The hierarchy of clusters can be represented as a tree structure called a dendrogram. Leaves of the dendrogram consist of one item as an individual cluster, while the root of the dendrogram contains all items belonging to one cluster. Internal nodes represent clusters formed by merging clusters of children, and the algorithm results in a sequence of groupings. The user then selects a "natural" clustering from this sequence.

## ILLUSTRATIVE EXAMPLE

To demonstrate the proposed methodology, an illustrative example has been developed. Quality inspection results (i.e., the number of inspected items and the number of repaired items) of eight types of products are detailed in Table 1. Each product type represents products with the same combination of design attributes. These data and the proposed approach are used to assess the product complexity indicator, score the product complexity indicator level, and cluster product complexity indicators.

### Step 1. Product Complexity Indicator

Following Eq. (3), theoretical distributions of product complexity indicators and median values of these distributions are derived as indicated in Table 2. When comparing the center of non-symmetric distributions, the median is the most appropriate statistical estimation. Accordingly, types 5 and 4 are the least and most complex products, respectively.

To visualize the theoretical beta distributions, a side-by-side box plot is developed by calculating the five-number summary (Min, Q1, Median, Q3, Max) and is illustrated in Figure 1. Medians of types 1, 2, 5, and 6 are approximately 2%, while product types 3, 4, 7, and 8 are approximately 4%. For paired types 1+2, 3+4, 5+6, and 7+8, each group has similar distribution spreads. Therefore, to account for the uncertainty of these distributions, the expected clusters should be product types 1+2, 3+4, 5+6, and 7+8.

***Step 2. Product Complexity Indicator Scoring***

By implementing the derived Hellinger distance equation for the two beta distributions, an $8 \times 8$ symmetric Hellinger distance matrix is constructed as shown in Eq. (8). This matrix corresponds to all the previously discussed properties of the Hellinger distance matrix and is used to perform the product clustering analysis.

$$M = \begin{bmatrix} 0.0000 & 0.0602 & 0.4100 & 0.4290 & 0.2109 & 0.2090 & 0.3900 & 0.3694 \\ 0.0602 & 0.0000 & 0.4023 & 0.4219 & 0.1566 & 0.1552 & 0.3989 & 0.3789 \\ 0.4100 & 0.4023 & 0.0000 & 0.0232 & 0.3737 & 0.3688 & 0.1604 & 0.1674 \\ 0.4290 & 0.4219 & 0.0232 & 0.0000 & 0.3937 & 0.3888 & 0.1703 & 0.1796 \\ 0.2109 & 0.1566 & 0.3737 & 0.3937 & 0.0000 & 0.0057 & 0.4100 & 0.3936 \\ 0.2090 & 0.1552 & 0.3688 & 0.3888 & 0.0057 & 0.0000 & 0.4046 & 0.3881 \\ 0.3900 & 0.3989 & 0.1604 & 0.1703 & 0.4100 & 0.4046 & 0.0000 & 0.0230 \\ 0.3694 & 0.3789 & 0.1674 & 0.1796 & 0.3936 & 0.3881 & 0.0230 & 0.0000 \end{bmatrix} \tag{8}$$

As per the medians from $P(0.5, Cplx_i | X_i)$ shown in Table 2, type 5 is characterized as the least complex. The product complexity indicator score can then be calculated through the recurrence relation as Eq. (6) by indexing the corresponding Hellinger distance matrix. The calculated complexity indicator scores, listed in Table 3, are within the range of 0 to 10. Several insights can be extracted from this result. For example, although the complexity indicator distributions of products 2 and 5 have similar medians, their $Cplx$ scores are quite different. This is primarily due to the variability in the spread of their complexity indicator distributions, which indicates that, even though products may possess similar median values, the product complexity may differ. This is also the reason for implementing a Hellinger distance to measure similarities among distributions.

***Step 3. Product Complexity Indicator Clustering***

To generate the dendrogram plot of the hierarchical clustering outcome, the statistical computing and graphics software, R ([https://www.r-project.org](https://www.r-project.org)), is used. Using the obtained Hellinger distance matrix and following the introduced agglomerative hierarchical clustering algorithm, 8 types of products are partitioned into clusters as shown in the dendrogram. To merge clusters of products, as opposed to individual products, the complete-linkage criterion is used to measure the distance between clusters. Products with small distance differences are grouped together. The heights (horizontal line) at which two clusters are merged represent the dissimilarity between two clusters in the data space. By specifying the expected number of clusters as four, types 1+2, 3+4, 5+6, 7+8 are grouped together (Figure 2) in a manner that is consistent with the visually-based prediction.

Given this illustrative example, the inherent mechanism of the proposed uncertain data clustering technique is comprehensively illustrated. The outcomes of all steps adequately verify the functionalities of this uncertain data clustering approach. In the next section, a practical case study will be conducted to validate the feasibility and applicability of the proposed methodology.

## CASE STUDY

Industrial construction is a construction method that involves large-scale use of offsite prefabrication and preassembly for building facilities, such as oil/gas production facilities and petroleum refineries. Pipe spool fabrication is crucial for the successful delivery of industrial projects. Pipe spool fabrication is heavily dependent on welding, which must be sampled and inspected to ensure that welding quality requirements are met. Typically, the difficulty (i.e., complexity) of pipe welds depends on various pipe attributes, such as nominal pipe size (NPS; the outside diameter of a pipe), schedule (wall thickness of a pipe), and material. In this section, an industrial pipe spool fabrication company in Edmonton, Canada, is studied to analyze pipe weld complexity using the proposed uncertain data clustering approach.

The case study is conducted following the data source identification, data adapter design, data analysis, and decision support procedures that are summarized in Figure 3. First, multiple data sources are investigated to extract useful information related to pipe weld quality performance and design attributes. Second, a data adapter is designed to efficiently connect data and map data into a single, tidy dataset. Then, the proposed uncertain data clustering approach is implemented to perform the product complexity analysis. Finally, main outputs are generated to produce new information and to support decision-making processes. All four procedures are performed using the statistical computing and graphics software, R (https://www.r-project.org).

### Data Source

The quality management and engineering design systems of the studied company were used to extract the non-destructive examination (NDE) inspection and pipe weld design attributes information, respectively. In this paper, only the inspection records of radiographic tests (RT) of butt welds were extracted from the quality management system. RT inspection results are tracked in three statuses for each pipe weld, namely: 0 – no inspection performed; 1 – inspected and passed; and 2 – inspected and failed. The engineering design system of the studied company stored pipe weld design attributes by the pipe format (NPS, schedule, material). For example, pipe [6, standard (STD), A] represents butt welds with NPS of 6, schedule of STD, and material A.

### Data Adapter

Since the multi-relational data required were dispersed across quality management and engineering design systems, a data adapter was required to collect useful information from various data sources into a single, centralized tidy dataset. In this case, a data adapter was also necessary to transform raw data, through data connection and data wrangling, into compatible, interpretable data formats. This is particularly important for data that are collected from a variety of sources or databases.

For data connection, the R package for Open Database Connectivity (RODBC) was used to connect to Structured Query Language (SQL) server of both the quality management and engineering design systems (Ripley et al. 2016). The dplyr/tidyr package was used to perform data wrangling tasks, including data reshaping, grouping, and combining (Wickham et al. 2017; Wickham et al. 2017). The completed dataset was transformed into a table format, where each variable was saved in its own column and each observation was saved in its own row. A sample for the centralized dataset is listed in Table 4. This dataset combines the pipe design attributes and quality inspection results.

For inferring the fraction nonconforming quality performance of each type of weld, all pipe welds were required to be grouped by pipe attributes (i.e., NPS, schedule, and material). Then, the data was summarized to count the total number of welds, inspected welds, and repaired welds for each type of pipe weld. A sample of the wrangled dataset is provided as Table 5. Each row represents the historical quality inspection information for a certain type of pipe weld. This table is then used to perform the complexity analysis as follows.

**Data Analysis**

Prior to performing the comprehensive product complexity indicator clustering analysis, the wrangled dataset was examined and relevant information was extracted and analyzed. A total of 224,298 welds comprised of 631 weld types were conducted over that last ten years. Figure 4 depicts detailed business percentages and cumulative business percentages of the top 35 types of pipe welds, which are those that have been selected for further analysis. As per the cumulative frequency graph shown in Figure 6, the top 35 types of pipe welds represented the most common pipe welding products and accounted for 80% of the company's business. Due to frame limitations of the graph, only the top 35 types of pipe welds were selected to perform the product complexity analysis.

The first step of the proposed methodology was applied to determine the product complexity indicator, with incorporated uncertainty, for each pipe weld type. Here, "Inspected Welds" and "Repaired Welds" from Table 5 were used to construct beta distributions as per Eq. (3). To be consistent with Figure 4, a side-by-side box plot, shown in Figure 5, was generated with the same sequence to visualize the distributions of the product complexity indicators. Notably, although some types followed similar distribution patterns, the product complexity indicators of these products varied considerably.

The complexity indicator distribution similarity between various groups of welding products was measured using the Hellinger distance metric. After obtaining the Hellinger distance matrix, welding products were scored based on the proposed scoring method. Also, the top 35 products were clustered into seven complexity groups, based on distribution similarity measurements, by using the agglomerative hierarchical clustering method (Figure 6). The name of each weld type is formatted as "Type ID.(NPS, schedule, material).$[Cplx\ Score]$" to include all design attributes and complexity indicator score information. Clusters are labelled from A to G based on the corresponding complexity level of that cluster, where Cluster A is the most complex group. The total business percentage of that cluster is also summarized and shown in Figure 6. The height at which two clusters are merged represents the dissimilarity between the two clusters in the data space. This type of information is expected to enable practitioners to better understand product complexity in a more informative and thorough manner.

**Validation**

To ensure that the proposed method is capable of reliably estimating product complexity, a systematic, expert evaluation-based validation was conducted. Eight welding operators, each with more than five-years working experience, were invited to evaluate the welding difficulty of selected weld types based on their own professional experience and knowledge. The authors excluded any "quality" related words from the evaluation description to eliminate any potential biases. The validation was conducted using the following protocol:

Step 1. Pick one weld type that accounts for the largest business percentage of each cluster (i.e., A to G). This will ensure that welding operators have sufficient experience with each of the chosen welds.

Step 2. Reshuffle the order of the selected weld types by changing the sequence in which they are presented to the welders.

Step 3. Invite welding operators to rank the welding difficulty based on their professional experience and knowledge using integers 1 to 7, where 7 represents the most complex weld type and 1 represents the least complex weld type.

Step 4. Once evaluations are collected, an average difficulty value is calculated for each type of pipe weld.

Step 5. Assign letter levels (i.e., A to G) to the sorted types of pipe welds based on the results.

Step 6. Compare the expert evaluation results with those obtained using the developed approach.

Table 6 demonstrates the detailed validation results. Although welding operator rankings were variable, the overall evaluated welding difficulty levels followed the same sequence that was obtained using the proposed framework. This validation outcome demonstrates the feasibility and applicability of the proposed product complexity approach and supports the hypothesis that product quality performance is associated with product complexity.

**Decision Support**

The management team from the studied company has confirmed that a data-driven decision support approach that enables the timely transformation of large datasets into useable knowledge is highly desirable in practice. To better support these practical needs, the product complexity analysis functionality has been incorporated into the previously developed simulation-based analytics framework (Ji and AbouRizk, 2017b). Once incorporated into the simulation-based analytics framework, product complexity levels, together with their detailed design information, can be targeted for management, design, and operation professionals who require this type of information in a timely manner.

Results of the proposed research, such as those obtained in the case study, are expected to enhance decision-making with the overall aim of improving the competitiveness and reputation of organizations within the industry. Three detailed perspectives identified through interviews with industrial professionals, by which the proposed method is expected to enhance decision-support processes, are described.

### Strategic Bidding

A better understanding of product complexity would assist practitioners in reducing uncertainty, which, in turn, could lead to improved cost performance. When bidding for new projects, practitioners may use the proposed framework to derive product complexity measurements in a relatively rapid manner that is conducive to the strict timelines associated with bid preparation. These measurements can then be used to allocate a contingency percentage that is more reflective of product complexity. For example, if a complex product is encompassing a majority of a bid, an organization should increase the contingency in their bid estimate. This would mitigate product complexity uncertainty and enhance the accuracy of bidding performance, thereby enhancing the company's profitability.

### Complexity-driven Production Planning

Previous research has developed a data-driven method to quantitatively identify exceptional operators for specific weld types (Ji and AbouRizk 2017b). Development and implementation of a complexity-driven production planning approach, which would allocate welding tasks to operators with high performance for each particular weld type, could directly improve overall welding quality performance and, consequently, reduce quality-induced rework cost and improve productivity. Such an automated production planning system, which incorporates both product complexity and detailed operator information, is expected to considerably increase the efficiency of industrial construction product prefabrication.

### Customized Training

Enhanced training programs are crucial for improving product complexity management processes and overall project performance. The proposed research transforms vast amounts of data into valuable knowledge in a simplified, interpretable, and informative format that efficiently improves practitioners' understanding of product complexity and reduces the time required to familiarize practitioners with this process. Once high-complexity products are identified, practitioners would be able to determine which operators are most proficient for each high-complex product. These operators should be invited to demonstrate their welding technique and share their professional knowledge for customized training purposes.

**CONCLUSION**

Product complexity is a predominant, yet often uncertain, factor that affects the success of construction project delivery. In this research, a novel uncertain data clustering approach was proposed to improve product complexity analysis by extracting hidden, intricate product complexity patterns from product quality performance measures. This approach contributes to the improved understanding of product complexity and, consequently, reduces the interpretation load for practitioners. Systematic procedures were developed for product complexity indicator determination, scoring, and clustering purposes. A pre-established product quality performance measurement, which incorporates uncertainty, is introduced as an indicator of product complexity. To the best of our knowledge, this is the first time that prefabricated construction product complexity is conceptually defined and quantitatively interpreted from the aspect of product quality performance.

The Hellinger distance is implemented to quantify the similarity of product complexity indicator distributions while considering uncertainty. In addition to providing a product complexity indicator score, the obtained Hellinger distance matrix is further utilized to perform the agglomerative hierarchical clustering method for intrinsically grouping products to achieve a better interpretation of product complexity. This novel Hellinger distance-based clustering approach is capable of clustering beta distributions and can be generalized and implemented for other types of uncertain data (i.e., probability distributions) clustering problems.

An industrial case study in Edmonton, Canada, was conducted to demonstrate the feasibility and applicability of the proposed uncertain data clustering approach. The achieved results indicate that the proposed method can appropriately cluster pipe weld types into homogeneous product complexity levels. Practitioners can implement this approach to enhance their product complexity management practices from the perspectives of (1) strategic bidding, (2) complexity-driven production planning, and (3) customized training.

Although this research proposes a novel approach to analyze construction product complexity, product complexity scores are, in fact, quality performance-based indicators of product complexity rather than a direct measure of complexity itself. In the future, additional indicators, such as productivity performance and safety performance, could be incorporated to measure product complexity in a more comprehensive and scientific way. Also, the authors would like to quantitatively correlate product complexity to design attributes and to forecast product complexity from product design information.

**DATA AVAILABILITY STATEMENT**

Data analyzed during the study were provided by a third party. Requests for data should be directed to the provider indicated in the Acknowledgements.

**ACKNOWLEDGEMENTS**

This research is funded by the NSERC Collaborative Research & Development (CRD) Grant (CRDPJ 492657). The authors would like to acknowledge Rob Reid, Doug McCarthy, Jason Davio, and Christian Jukna at Falcon Fabricators and Modular Builders Ltd. for sharing their knowledge and expertise of industrial pipe welding complexity and quality management.

**APPENDIX 1 –**
**Proof of Hellinger distance for two beta distributions**

Let $P = \{p_i\}$ i $\in$ [n] and $Q = \{q_i\}$ i $\in$ [n] be two probability distributions supported on [n].

The Hellinger distance between two probability distributions is defined by:

$$H(P, Q) = \frac{1}{\sqrt{2}} \left\| \sqrt{P} - \sqrt{Q} \right\|_2$$

$$= \frac{1}{\sqrt{2}} \sqrt{\sum_i^n (\sqrt{p_i} - \sqrt{q_i})^2}$$

$$= \frac{1}{\sqrt{2}} \sqrt{\int (\sqrt{p_i} - \sqrt{q_i})^2}$$

Let $X_t$ and $Y_t$ be two independent beta probability distributions where:

$$X_t \sim Beta(a_1, b_1) \quad Y_t \sim Beta(a_2, b_2)$$

$$H(X_t, Y_t) = \frac{1}{\sqrt{2}} \sqrt{\int (\sqrt{X_t} - \sqrt{Y_t})^2 dt}$$

$$H^2(X_t, Y_t) = \frac{1}{2} \int (\sqrt{X_t} - \sqrt{Y_t})^2 dt$$

$$H^2(X_t, Y_t) = \frac{1}{2} \left[ \int \sqrt{X_t}^2 dt - 2 \int \sqrt{X_t}\sqrt{Y_t}\, dt + \int \sqrt{Y_t}^2 dt \right]$$

$$H^2(X_t, Y_t) = \frac{1}{2} \left[ \int X_t\, dt - 2 \int \sqrt{X_t Y_t}\, dt + \int Y_t\, dt \right]$$

The integral of a probability density over its domain equals to 1.

$$H^2(X_t, Y_t) = \frac{1}{2} \left[ 1 - 2 \int \sqrt{X_t Y_t} dt + 1 \right]$$

$$H^2(X_t, Y_t) = 1 - \int \sqrt{X_t Y_t}\, dt$$

Probability density function of beta distribution is defined as:

$$f_t = \frac{(t-u)^{a-1}(l-t)^{b-1}}{Beta(a,b)(u-l)^{a+b-1}}$$

Where $a, b$ are shape parameters and $u, l$ are upper and lower boundaries.

Specify lower and upper boundaries, respectively, as 0 and 1, probability density function of beta distribution becomes:

$$f_t = \frac{t^{a-1}(1-t)^{b-1}}{Beta(a,b)}$$

Beta function is defined as:

$$Beta(a,b) = \int_0^1 t^{a-1}(1-t)^{b-1} \, dt$$

Based on the function of beta distribution, the squared Hellinger distance can be written as:

$$H^2(X_t, Y_t) = 1 - \int_0^1 \sqrt{\frac{t^{a_1-1}(1-t)^{b_1-1}}{Beta(a_1,b_1)} \times \frac{t^{a_2-1}(1-t)^{b_2-1}}{Beta(a_2,b_2)}} \, dt$$

$$H^2(X_t, Y_t) = 1 - \frac{1}{\sqrt{Beta(a_1,b_1) \times Beta(a_2,b_2)}} \int_0^1 \sqrt{t^{a_1-1}(1-t)^{b_1-1} \times t^{a_2-1}(1-t)^{b_2-1}} \, dt$$

$$H^2(X_t, Y_t) = 1 - \frac{1}{\sqrt{Beta(a_1,b_1) \times Beta(a_2,b_2)}} \int_0^1 \sqrt{t^{a_1+a_2-2} \times (1-t)^{b_1+b_2-2}} \, dt$$

$$H^2(X_t, Y_t) = 1 - \frac{1}{\sqrt{Beta(a_1,b_1) \times Beta(a_2,b_2)}} \int_0^1 \sqrt{t^{2(\frac{a_1+a_2}{2}-1)} \times (1-t)^{2(\frac{b_1+b_2}{2}-1)}} \, dt$$

$$H^2(X_t, Y_t) = 1 - \frac{1}{\sqrt{Beta(a_1,b_1) \times Beta(a_2,b_2)}} \int_0^1 t^{(\frac{a_1+a_2}{2}-1)} \times (1-t)^{(\frac{b_1+b_2}{2}-1)} \, dt$$

$$H^2(X_t, Y_t) = 1 - \frac{Beta(\frac{a_1+a_2}{2}, \frac{b_1+b_2}{2})}{\sqrt{Beta(a_1,b_1) \times Beta(a_1,b_2)}}$$

$$H(X_t, Y_t) = \sqrt{1 - \frac{Beta(\frac{a_1+a_2}{2}, \frac{b_1+b_2}{2})}{\sqrt{Beta(a_1,b_1) \times Beta(a_2,b_2)}}}$$

**REFERENCES**

Antani, K. R. (2014). "A study of the effects of manufacturing complexity on product quality in mixed-model automotive assembly." Doctor of Philosophy (PhD), CLEMSON UNIVERSITY.

Baccarini, D. (1996). "The concept of project complexity—a review." *International Journal of Project Management*, 14(4), 201-204.

Baldwin, C. Y., and Clark, K. B. (2000). *Design rules: The power of modularity*, MIT press.

Galvin, P., and Morkel, A. (2001). "The effect of product modularity on industry structure: the case of the world bicycle industry." *Industry and Innovation*, 8(1), 31.

Han, J., Pei, J., and Kamber, M. (2011). *Data mining: concepts and techniques*, Elsevier.

Hellinger, E. (1909). "Neue Begründung der Theorie quadratischer Formen von unendlichvielen Veränderlichen." *Journal für die reine und angewandte Mathematik*, 136, 210-271.

Ji, W., and AbouRizk, S. M. (2017a). "Credible estimation for fraction nonconforming: analytical and numerical solutions." *Automation in Construction*, 83, 56-67.

Ji, W., and AbouRizk, S. M. (2017b). "Simulation-based analytics for quality control decision support: a pipe welding case study." *Journal of Computing in Civil Engineering*, https://doi.org/10.1061/(ASCE)CP.1943-5487.0000755. [Accepted: Oct. 23, 2017].

Jiang, B., Pei, J., Tao, Y., and Lin, X. (2013). "Clustering uncertain data based on probability distribution similarity." *IEEE Transactions on Knowledge and Data Engineering*, 25(4), 751-763.

Liese, F., and Vajda, I. (2006). "On divergences and informations in statistics and information theory." *IEEE Transactions on Information Theory*, 52(10), 4394-4412.

Luo, L., He, Q., Jaselskis, E. J., and Xie, J. (2017). "Construction Project Complexity: Research Trends and Implications." *Journal of Construction Engineering and Management*, https://doi.org/10.1061/(ASCE)CO.1943-7862.0001306, 04017019.

Montgomery, D. C. (2007). *Introduction to statistical quality control*, John Wiley & Sons.

Novak, S., and Eppinger, S. D. (2001). "Sourcing by design: Product complexity and the supply chain." *Management science*, 47(1), 189-204.

Pei, J., Jiang, B., Lin, X., and Yuan, Y. (2007). "Probabilistic skylines on uncertain data." *Proc., Proceedings of the 33rd international conference on Very large data bases*, VLDB Endowment, 15-26.

Ripley, B., Lapsley, M., and Ripley, M. B. (2016). "Package 'RODBC'", <https://cran.r-project.org/web/packages/RODBC/index.html > (Sept. 19, 2017).

Schubert, E., Koos, A., Emrich, T., Züfle, A., Schmid, K. A., and Zimek, A. (2015). "A framework for clustering uncertain data." *Proceedings of the VLDB Endowment*, 8(12), 1976-1979.

Senescu, R. R., Aranda-Mena, G., and Haymaker, J. R. (2012). "Relationships between project complexity and communication." *Journal of Management in Engineering*, https://doi.org/10.1061/(ASCE)ME.1943-5479.0000121, 183-197.

Wickham, H., Henry, L., and RStudio (2017). "tidyr: Easily Tidy Data with spread () and gather () Functions.", < https://cran.r-project.org/web/packages/tidyr/tidyr.pdf> (Sept. 14, 2017).

Wickham, H., Francois, R., Henry, L., Müller, K., and RStudio (2017). "dplyr: A grammar of data manipulation.", < https://cran.r-project.org/web/packages/dplyr/dplyr.pdf > (Sept. 14, 2017).

Williams, T. M. (1999). "The need for new paradigms for complex projects." *International journal of project management*, 17(5), 269-273.

Züfle, A., Emrich, T., Schmid, K. A., Mamoulis, N., Zimek, A., and Renz, M. (2014). "Representative clustering of uncertain data." *Proc., Proceedings of the 20th ACM SIGKDD international conference on Knowledge discovery and data mining*, ACM, 243-252.

**FIGURE LIST**

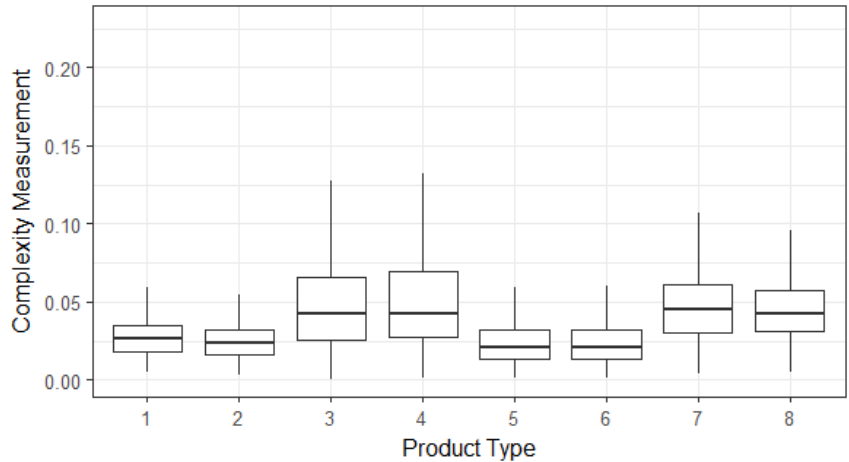

**Figure 1.** Side-by-side boxplot for eight types of product complexity measurements.

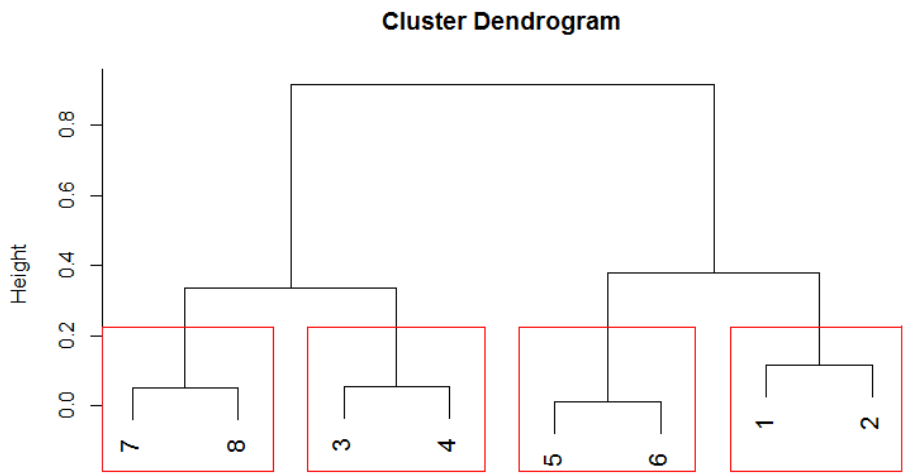

**Figure 2.** Cluster dendrogram of the illustrative example.

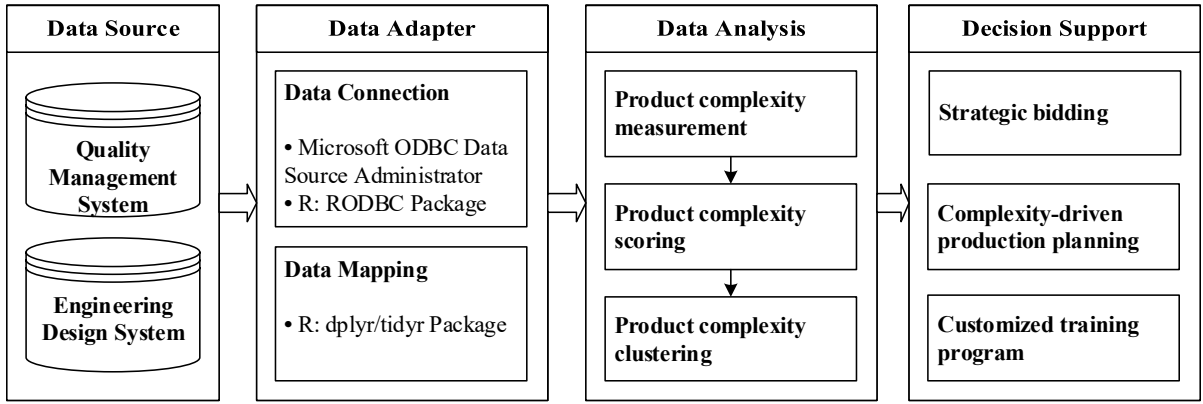

**Figure 3.** Workflow of the case study.

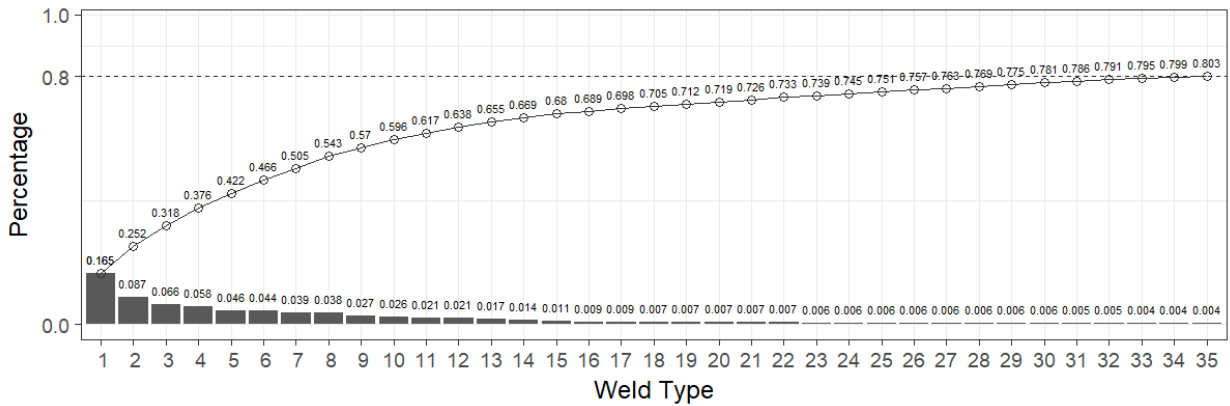

**Figure 4.** Cumulative percentage of the top 35 weld types.

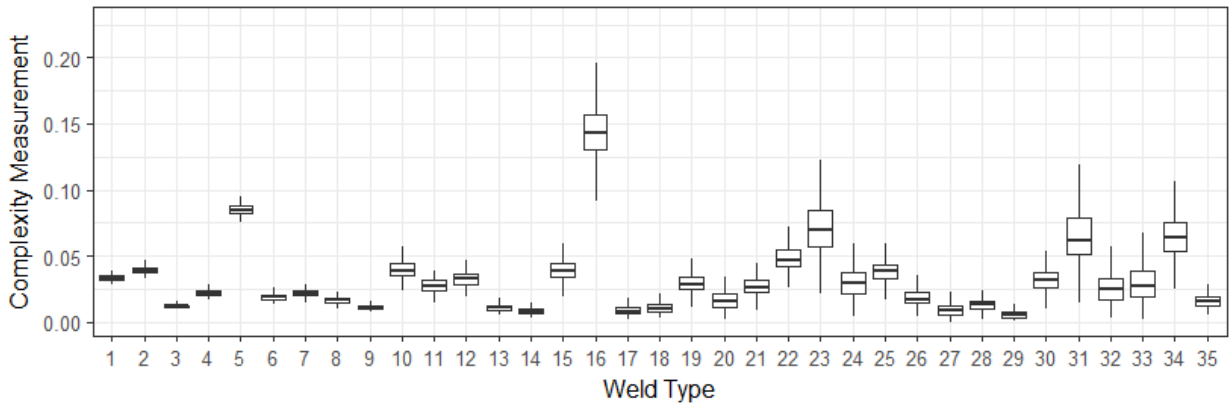

**Figure 5.** Complexity measurements of the top 35 weld types.

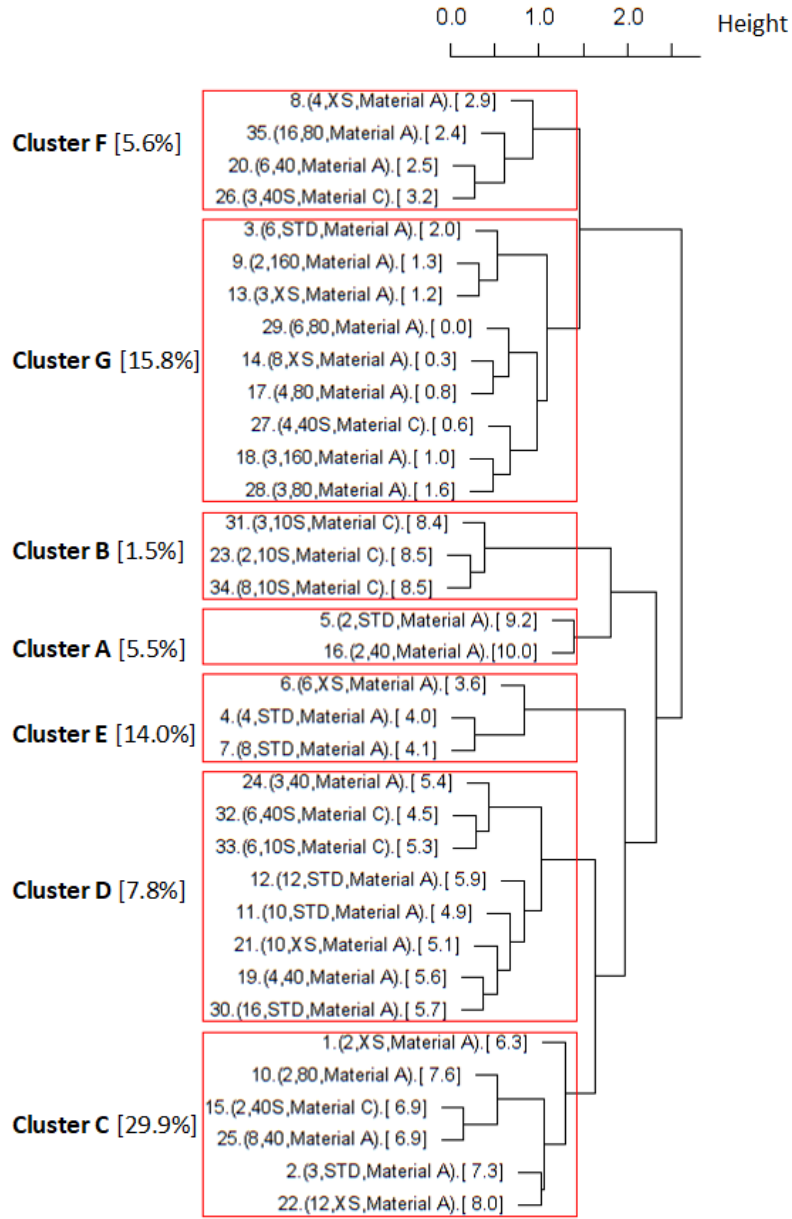

**Figure 6.** Complexity clustering dendrogram of the top 35 weld types.

**Table 1.** Quality inspection results of eight types of products.

| Product Type | Number of Inspected Items | Number of Repaired Items |
|---|---|---|
| 1 | 200 | 5 |
| 2 | 170 | 4 |
| 3 | 50 | 2 |
| 4 | 48 | 2 |
| 5 | 100 | 2 |
| 6 | 99 | 2 |
| 7 | 98 | 4 |
| 8 | 101 | 4 |

**Table 2.** Product complexity distributions and median values.

| Product Type | $P(Cplx\|X)$ | $P(0.5, Cplx_i\|X_i)$ |
|---|---|---|
| 1 | Beta (5.5, 195.5) | 0.0258 |
| 2 | Beta (4.5, 166.5) | 0.0245 |
| 3 | Beta (2.5, 48.5) | 0.0432 |
| 4 | Beta (2.5, 46.5) | 0.0450 |
| 5 | Beta (2.5, 98.5) | 0.0217 |
| 6 | Beta (2.5, 97.5) | 0.0219 |
| 7 | Beta (4.5, 94.5) | 0.0424 |
| 8 | Beta (4.5, 97.5) | 0.0412 |

**Table 3.** Complexity scores for eight types of products.

| Product Type | $Cplx\ Score$ | $P(0.5, Cplx_i\|X_i)$ |
|---|---|---|
| 1 | 2.8 | 0.0258 |
| 2 | 2.0 | 0.0245 |
| 3 | 9.7 | 0.0432 |
| 4 | 10.0 | 0.0450 |
| 5 | 0.0 | 0.0217 |
| 6 | 0.1 | 0.0219 |
| 7 | 7.7 | 0.0424 |
| 8 | 7.4 | 0.0412 |

**Table 4.** A data sample for the centralized dataset.

| Weld Type | NPS | Schedule | Material | Inspection Result |
|---|---|---|---|---|
| 1 | 10 | 40S | B | 1 |
| 2 | 2 | 40 | C | 0 |
| 3 | 6 | XS | D | 2 |
| … | … | … | … | … |

**Table 5.** Wrangled dataset of the top 35 weld types.

| Weld Type | NPS | Schedule | Material | Total Welds | Inspected Welds | Repaired Welds |
|---|---|---|---|---|---|---|
| 1 | 2 | XS | Material A | 37059 | 7475 | 249 |
| 2 | 3 | STD | Material A | 19464 | 4495 | 173 |
| 3 | 6 | STD | Material A | 14866 | 3518 | 43 |
| 4 | 4 | STD | Material A | 13020 | 3078 | 66 |
| 5 | 2 | STD | Material A | 10304 | 4722 | 400 |
| 6 | 6 | XS | Material A | 9916 | 3705 | 70 |
| 7 | 8 | STD | Material A | 8722 | 2302 | 51 |
| 8 | 4 | XS | Material A | 8601 | 1774 | 28 |
| 9 | 2 | 160 | Material A | 6044 | 2302 | 26 |
| 10 | 2 | 80 | Material A | 5854 | 1055 | 41 |
| 11 | 10 | STD | Material A | 4822 | 1131 | 30 |
| 12 | 12 | STD | Material A | 4728 | 1069 | 34 |
| 13 | 3 | XS | Material A | 3733 | 1484 | 16 |
| 14 | 8 | XS | Material A | 3193 | 1318 | 10 |
| 15 | 2 | 40S | Material C | 2431 | 555 | 21 |
| 16 | 2 | 40 | Material A | 2088 | 271 | 38 |
| 17 | 4 | 80 | Material A | 2056 | 638 | 5 |
| 18 | 3 | 160 | Material A | 1676 | 510 | 5 |
| 19 | 4 | 40 | Material A | 1550 | 592 | 17 |
| 20 | 6 | 40 | Material A | 1673 | 333 | 5 |
| 21 | 10 | XS | Material A | 1676 | 529 | 14 |
| 22 | 12 | XS | Material A | 1652 | 666 | 31 |
| 23 | 2 | 10S | Material C | 1261 | 175 | 12 |
| 24 | 3 | 40 | Material A | 1358 | 217 | 6 |
| 25 | 8 | 40 | Material A | 1413 | 452 | 17 |
| 26 | 3 | 40S | Material C | 1441 | 364 | 6 |
| 27 | 4 | 40S | Material C | 1253 | 271 | 2 |
| 28 | 3 | 80 | Material A | 1436 | 512 | 6 |
| 29 | 6 | 80 | Material A | 1407 | 572 | 3 |
| 30 | 16 | STD | Material A | 1406 | 422 | 13 |
| 31 | 3 | 10S | Material C | 1117 | 149 | 9 |
| 32 | 6 | 40S | Material C | 1128 | 171 | 4 |
| 33 | 6 | 10S | Material C | 912 | 154 | 4 |
| 34 | 8 | 10S | Material C | 912 | 204 | 13 |
| 35 | 16 | 80 | Material A | 961 | 634 | 9 |

**Table 6.** Detailed validation results.

| Clustered Complexity | Weld Type | Design Attributes | Welding Difficulty Evaluation | | | | | | | | | |
|---|---|---|---|---|---|---|---|---|---|---|---|---|
| | | | Welder 1 | Welder 2 | Welder 3 | Welder 4 | Welder 5 | Welder 6 | Welder 7 | Welder 8 | Average | Letter Level |
| A | 5 | (2, STD, Material A) | 7 | 7 | 7 | 7 | 7 | 7 | 7 | 7 | 7.0 | A |
| B | 23 | (2, 10S, Material C) | 6 | 6 | 6 | 6 | 6 | 6 | 6 | 6 | 6.0 | B |
| C | 1 | (2, XS, Material A) | 5 | 4 | 5 | 5 | 5 | 5 | 3 | 5 | 4.6 | C |
| D | 11 | (10, STD, Material A) | 3 | 5 | 4 | 3 | 4 | 3 | 5 | 3 | 3.8 | D |
| E | 4 | (4, STD, Material A) | 4 | 3 | 3 | 4 | 3 | 4 | 4 | 4 | 3.6 | E |
| F | 8 | (4, XS, Material A) | 1 | 2 | 2 | 2 | 1 | 2 | 2 | 1 | 1.6 | F |
| G | 3 | (6, STD, Material A) | 2 | 1 | 1 | 1 | 2 | 1 | 1 | 2 | 1.4 | G |